\documentclass[aps,pre,twocolumn,superscriptaddress,floatfix,10pt]{revtex4-1}
\usepackage{graphicx}
\usepackage{dcolumn}
\usepackage{bm}
\usepackage{latexsym}
\usepackage{amssymb}
\usepackage{amsmath}

\begin{document}

\title{First-passage processes on a filamentous track in a dense traffic: Optimizing diffusive search for a target in crowding conditions}
\author{Soumendu Ghosh}
\affiliation{Department of Physics, Indian Institute of Technology
  Kanpur, 208016, India}
\author{Bhavya Mishra}
\affiliation{Department of Physics, Indian Institute of Technology
  Kanpur, 208016, India}
\author{Anatoly B. Kolomeisky}
\affiliation{Department of Chemistry, Department of Chemical and Biomolecular Engineering and Center for Theoretical Biological Physics, Rice University, Houston, TX 77005, USA}
  \author{Debashish Chowdhury} 
\email[Corresponding author; E-mail:]{debch@iitk.ac.in}
\affiliation{Department of Physics, Indian Institute of Technology
  Kanpur, 208016, India}

\begin{abstract}

Several important biological processes are initiated by the binding of a protein to a specific site on the DNA. The strategy adopted by a protein, called transcription factor (TF), for searching its specific binding site on the DNA has been investigated over several decades. In recent times the effects obstacles, like DNA-binding proteins, on the search by TF has begun to receive attention. RNA polymerase (RNAP) motors collectively move along a segment of the DNA during a genomic process called transcription. This RNAP traffic is bound to affect the diffusive scanning of the same segment of the DNA by a TF searching for its binding site. Motivated by this phenomenon, here we develop a kinetic model where a `particle', that represents a TF, searches for a specific site on a one-dimensional lattice. On the same lattice another species of particles, each representing a RNAP, hop from left to right exactly as in a totally asymmetric simple exclusion process (TASEP) which forbids simultaneous occupation of any site by more than one particle, irrespective of their identities. Although the TF is allowed to attach to or detach from any lattice site, the RNAPs can attach only to the first site at the left edge and detach from only the last site on the right edge of the lattice. We formulate the search as a {\it first-passage} process; the time taken to reach the target site {\it for the first time}, starting from a well defined initial state, is the search time. By approximate analytical calculations and Monte Carlo (MC) computer simulations, we calculate the mean search time. We show that RNAP traffic rectifies the diffusive motion of TF to that of a Brownian ratchet, and the mean time of successful search can be even shorter than that required in the absence of RNAP traffic. Moreover, we show that there is an optimal rate of detachment that corresponds to the shortest mean search time. 

\end{abstract}

\maketitle 

\section{introduction}

To initiate most of the biological processes in living cells, protein molecules, the workhorses of the cell, have to bind to specific sites on nucleic acid molecules \cite{alberts,bressloff}. One of the common examples, which is also crucially important for the functioning of a cell, is the binding of transcription factor (TF) proteins to a specific sequence site on the DNA. Protein has to `search' for this binding site, but such a `search' carried out by a TF is neither guided by any external cues nor does it benefit from past experience because of the absence of any memory. Instead, this search is believed to be a random (stochastic) process \cite{bressloff,halford04,mirny09,kolomeisky11}. 

A one-dimensional diffusive scanning of the DNA strand by the TF protein constitutes one mode of search that is combined with other possible modes, including dissociation and diffusion in  the bulk solution, re-association back to DNA and other possibilities \cite{bressloff,halford04,mirny09,kolomeisky11}. For proper biological function, the search strategy should not only be fast but must also rule out the possibility of erroneous recognition of any other site as the intended target site of binding. Enormous progress have been made in the last few decades in understanding the strategies evolved by nature by combining various possible modes of search by a TF that optimize the opposite demands of speed and accuracy of search in a cell \cite{halford04,mirny09,kolomeisky11,veksler13,tafvizi11,sheinman12,kolomeisky12,kolomeisky16}. 

In live cells, the search by a TF is made difficult by the fact that the target binding sites are usually located in an extremely crowded environment. The molecules surrounding the DNA strand reduce the accessibility of the target site while those bound to DNA create a steric hindrance against scanning of the DNA chain \cite{shvets16,shvets15b}. Since a dissociation of a TF from DNA, followed by a subsequent re-attachment elsewhere on DNA, is an integral part of its search strategy, a TF does not remain permanently obstructed by any DNA-bound molecule. Nevertheless, the blockages created by such DNA-bound particles against the diffusive search by TFs can have significant non-trivial effects on dynamics of the search process. This phenomenon has already attracted the attention of theorists in recent years \cite{veksler13,shvets16,shvets15b}. 

Often what makes the search problem even more challenging is that many DNA-bound molecules are themselves mobile so that, during the diffusive scanning of the DNA  chain for its target binding site, the TF encounters the mobile obstacles either co-directionally or head-on. For example, RNA polymerases (RNAPs), for which a segment of DNA serves as the  template for the synthesis (polymerization) of a specific molecular species of RNA, use the template DNA strand also as a track for its motor-like \cite{chowdhury13,kolomeisky15} walk in a directed manner \cite{alberts,buc}. The process of synthesis of RNA, as directed by a DNA template, is called transcription (of a gene) \cite{alberts}. The traffic of RNAPs \cite{tripathi08,klumpp08,klumpp11,sahoo11,ohta11,wang14,belitsky18}, engaged in transcription, would act as an oncoming stream of mobile roadblocks against a TF that simultaneously searches for its specific binding site on the same segment of DNA. Besides, at any given time, many TFs can search for the same segment of DNA for their respective binding sites and, therefore, their interactions are also likely to affect their individual efficiency of search. 

To our knowledge, no theoretical model has been developed so far to study the effects of DNA-bound mobile molecules on the diffusive search by TFs on the same segment of DNA.  In this paper we develop a kinetic model motivated by the search of specific binding sites by a single TF as well as that by several TFs simultaneously over a segment of DNA that is also undergoing transcription by a traffic of RNAP motors. This minimal model is not intended to comprehensively describe  {\it in-vitro} or {\it in-vivo}  experimental observations with any specific cell or organism. Instead, the main aim of our biologically motivated kinetic model  is to clarify a complex molecular picture and reveal interesting physics that such stochastic systems are likely to exhibit. 

In our model, we represent the TFs and RNAPs by two distinct species of particles. Although not all the features of TFs and RNAPs are taken into account in our analysis, these are still called TF and RNAP for the sake of simplicity of terminology. The  RNAPs hop forward uni-directionally on a discrete lattice of DNA sites while the TFs perform unbiased random walk on the same lattice. There is a pair of specially designated sites for the entry and exit of the RNAPs; these particles cannot attach to, or detach from, the lattice at any other site in between. In contrast, the TFs can attach to-, and detach from, any site on the lattice. The model captures the key features of RNAP traffic by a totally asymmetric simple exclusion process (TASEP) \cite{derrida98,schutz00,Schadschneider10,mallick15} which is one of the simplest models of collective stochastic movement of interacting self-propelled particles on a one-dimensional lattice. Since none of the sites can be occupied simultaneously by more than one particle, irrespective of the identity of the occupant, mutual exclusion is the only intra-species (RNAP-RNAP and TF-TF) as well as inter-species (RNAP-TF) interaction in our model.

The process of target search by a TF is treated in our theoretical framework as a {\it first-passage} process; starting from a given unique initial state, the time required for the completion of each successful search is a {\it first-passage time} (FPT) \cite{bressloff,redner,redner14}. Since the search process is stochastic, the search time is a random variable whose probability distribution is one of the main quantities of our interest here. The goal of this paper is to investigate the effects of intra-species and inter-species interactions among the particles on the search dynamics of the TFs. Our theoretical studies of the model are based on (approximate) analytical calculations and computer simulations. Among the various phenomena that we observe the following are most notable: (i) Over a range of parameter values, the search by the TF follows a mechanism that is similar to a Brownian ratchet; (ii) For given values of all the other parameters, there is an optimum rate of detachment of TF from the lattice at which its mean search time attains a minimum. We explain the underlying physical principles that give rise to these interesting phenomena.

\section{model}

\begin{figure}[h]
  \includegraphics[angle=0,width=1.0\columnwidth]{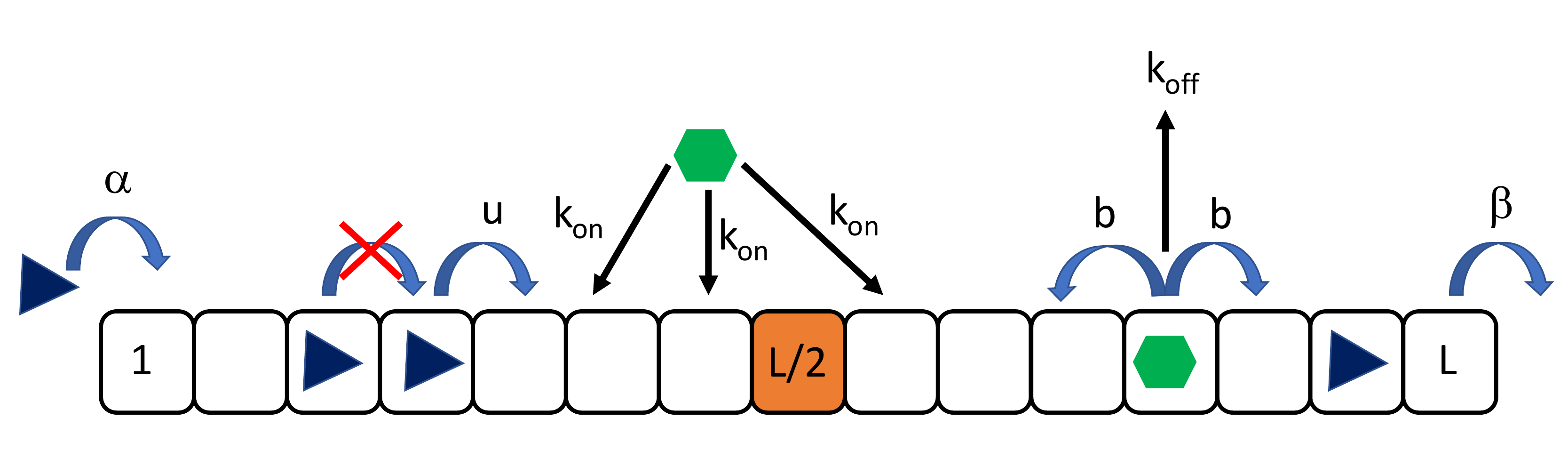}
  \caption{The schematic representation of the model. The model consists a one dimensional lattice and multiple particles of two different species. The first species particles represent TFs, whereas, second species particles represent RNAPs. A TF can attach at any site throughout the lattice with the rate $ k_{on} $ as well as it can detach from any site with rate $ k_{off} $. Inside the lattice a TF can hop in both the forward and backward directions with identical rates $b$. Unlike TF, a RNAP can attach only at site $ i=1 $, with rate $ \alpha $ and it can detach only from site $ i=L $, with rate $ \beta $. Inside the lattice, a RNAP can jump only in the forward direction, with the rate $u$. All the particles follow the exclusion principle, i.e., no two particles can occupy the same site simultaneously.}
  \label{fig-model}
\end{figure}

The kinetics of the model is shown schematically in Fig. \ref{fig-model}. The model consists of a one dimensional lattice with equispaced lattice sites that are labeled by the integer index $i$ ($1 \leq i \leq L$).  Thus, the length of the lattice is $L$ in the units of lattice spacing. Throughout this paper we assume that there is a single TF searching for a specific site located on the one-dimensional lattice. A TF can attach to an arbitrary lattice site, with a rate $k_{on} $, provided that this site is empty. Once attached to the lattice, the TF can hop forward or backward till it detaches from the lattice. The TF that is already attached to the lattice can detach from it, with a detachment rate $k_{off}$. Inside the bulk of the system, i.e., at sites $2 \leq i \leq L-1$, the TF can hop both forward or backward, with a rate $b$. This unbiased random walk (RW) of the particle captures the one-dimensional diffusion of the TF on the DNA chain. At the edges of the lattice, i.e., at $i=1$ and $i=L$, the TF can hop, with the rate $b$,  only in the forward and backward directions, respectively. 

The position of the TF is marked by an integer index $n$, where $n$ is allowed to vary over the range $0 \leq n \leq L$. The positions $n=1,2, \dots, L$ of the TF coincide with the lattice sites $i=1,2, \dots, L$, whereas the position $n=0$ indicates being in the solution (i.e. the medium in which the lattice exists).  Unlike the TFs, a RNAP can attach to the lattice, with rate $\alpha$, only at the site $i=1$, provided that this site is not already occupied by another RNAP or TF. Once attached to the lattice, the RNAP can hop only in the forward direction, with a rate $u$, while respecting the exclusion principle at each step.  The RNAP continues its forward hopping at the given rate till it reaches the last lattice site labeled by $i=L$ from where it is allowed to detach from the lattice with a rate $\beta$. Throughout the paper, we assume that the TASEP of RNAPs attain its non-equilibrium steady state, characterized by a time-independent flux, before the search by TFs begin.


\section{Results}  

Since it has not been possible to obtain a single analytical expression that would be valid in all regimes of the parameters of our model, we derive these expressions in two different limits, namely low- and high-rates of detachments from the lattice. We also check the accuracy of the analytical expressions derived in these two limits by comparing those with the corresponding numerical data obtained from Monte Carlo (MC) simulations of the model.

\subsection{Vanishing rate of detachment: insight from the extreme limit}

In order to get insight into the effects of the traffic of the RNAPs  on the kinetics of the TF, let us begin with the extreme limiting case of $k_{off}=0$. If the TF is assumed to begin at $t=0$ in a state where it is already attached to the lattice at a site $n \neq m$, it remains attached at all times $t > 0$. The kinetics is still interesting and depends on the dynamical phase that the RNAPs are expected to exhibit for the chosen set of values of the parameters in the complete absence of any TF. 

\begin{figure}[h]
  \includegraphics[angle=0,width=1.0\columnwidth]{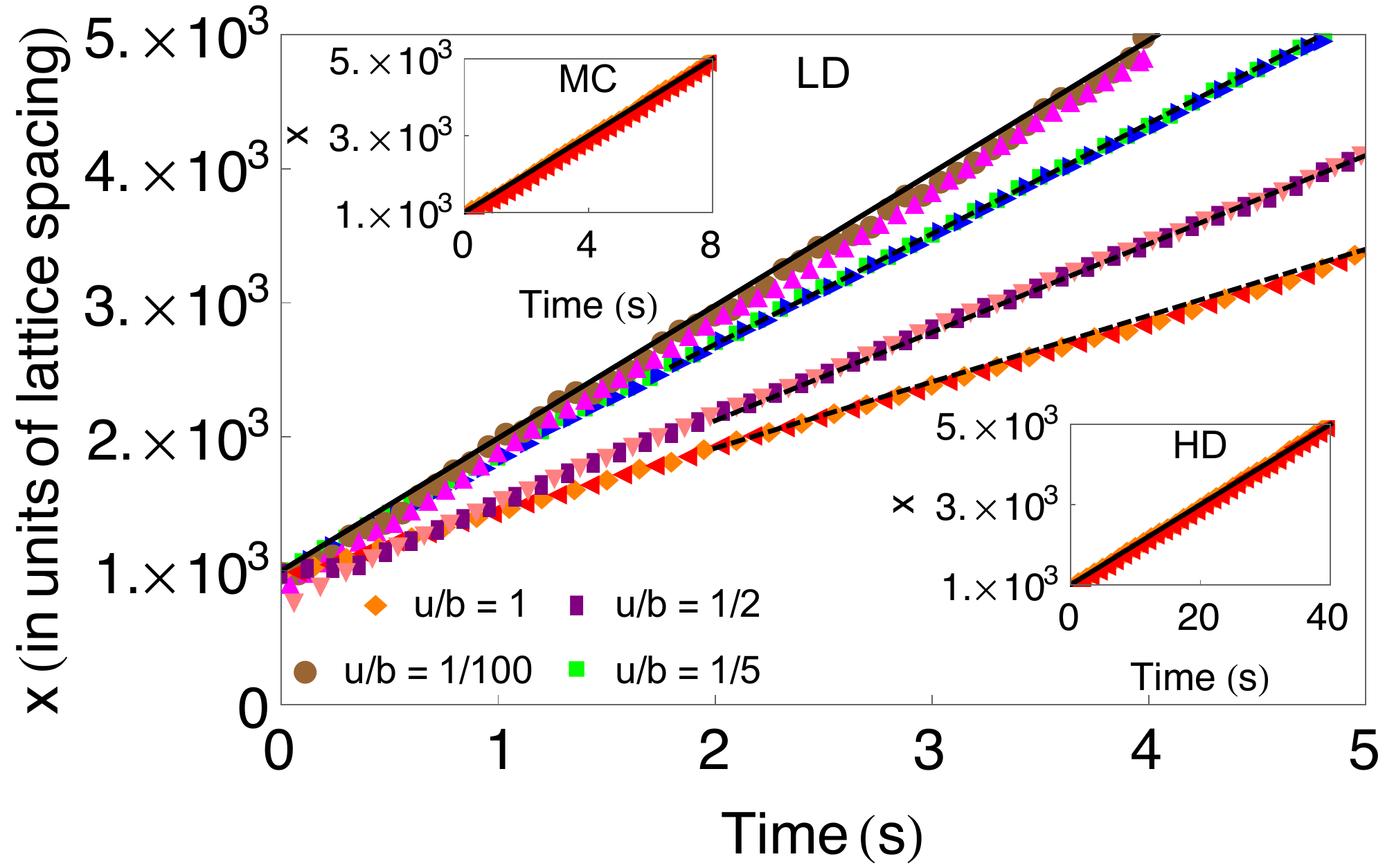}
  \caption{Position ($x$) of the TF and the RNAP immediately on its left are plotted against time ($t$) in the LD phase of the RNAP traffic for four different values of the ratio $u/b$. All the four triangles correspond to the RNAP whereas the remaining four symbols correspond to the TF. The corresponding data in the HD phase and MC phase of the RNAP traffic are  plotted in the insets; the data are practically independent of the value of the ratio $u/b$. The solid black lines indicate average velocities of the RNAPs while the black dashed lines correspond to $v_{eff}$.}
  \label{fig-xt0koff}
\end{figure}

Intuitively, it is obvious that the RNAPs prevent the backward (left-ward according to the Fig. \ref{fig-model}) steppings of the TF. Therefore, the TF behaves as a Brownian ratchet and exhibits a forward-directed (i.e., towards right) motion. Consequently, the TF can never reach the target site in this extreme limit if initially $n > m$, i.e., the searcher is located initially on the right side of the target. 

In contrast, if initially $n < m$, i.e., the search begins from a site on the left of the target site, the searcher TF would certainly hit the target after some time during which it is closely followed by a RNAP that rectifies the Brownian motion of the searcher. 
In Fig.\ref{fig-xt0koff} we plot the position of the TF and the closest RNAP following it , both as functions of time. The asymptotic linear increase of the position $x$ with time $t$ establishes that the TF moves, effectively, ballistically instead of its natural diffusive search, because of the rectification of the backward steps by the RNAP following it from behind. 

Suppose the parameters $\alpha$, $\beta$ and $u$ are such that the RNAPs  would be in the LD phase in the absence of any TF. In this case the mean search time is 
\begin{equation}
<t_s> \simeq (m-n)/v_{eff} 
\end{equation}
where the effective average velocity of the TF should be 
\begin{equation}
v_{eff} = \biggl(\frac{b}{b+u}\biggr) v = \biggl(\frac{1}{1 + (u/b)}\biggr) v
\label{eq-v0koff}
\end{equation} 
with 
\begin{equation}
v = u \biggl(1- \frac{\alpha}{u}\biggr)
\label{eq-LDavVel}
\end{equation}
being the corresponding average velocity that the RNAP would have in the LD phase in the absence of any TF. Indeed, the RNAPs do achieve the average velocity (\ref{eq-LDavVel}) if $u/b$ is sufficiently small (see Fig.\ref{fig-xt0koff}) because the faster moving TF vacates the site in front of the following RNAP sooner thereby leaving the RNAP practically unaffected by the TF. The formula (\ref{eq-v0koff}) is in excellent agreement with the simulation data plotted in Fig.\ref{fig-xt0koff} where we have used $\alpha=  10$ s$^{-1}$, $\beta=10^3$ s$^{-1}$, $u = 10^3$ s$^{-1}$ and, hence, $v = 990$ s$^{-1}$. Note that in the HD phase the mean search time is $<t_s> = (m-n)/v$, where $v=\beta$, irrespective of the magnitude of $b/u$, because the gap between successive RNAPs is so small that the TF is essentially dragged by the flow of the RNAPs with the same velocity as that of the RNAPs.

For the $x-t$ plots in both the insets of Fig.\ref{fig-xt0koff} we have used the parameters $u= b = 10^{3}$ s$^{-1}$. The other parameters, namely, $\alpha$ and $\beta$ were selected so as to attend the desired phase of the RNAPs in the absence of TF. For attaining the HD phase we chose $\alpha =1000$ s$^{-1}$ and  $\beta=100$ s$^{-1}$. In contrast, for attaining the MC phase we selected $\alpha=1000$ s$^{-1}$ and  $\beta=1000$ s$^{-1}$. So, in the steady state, the average velocity of the RNAPs would be $v = u (\beta/u) = \beta = 10^{2}$ s$^{-1}$. The slope of the straight line in the corresponding inset of Fig.\ref{fig-xt0koff} is, indeed, $10^{2}$ s$^{-1}$. Similarly, the average velocity of the RNAPs in the MC phase is $u (1 - \rho) = u/2$, as $\rho = 1/2$. Moreover, since we have taken $u = 1000$ s$^{-1}$, the average velocity of the RNAPs in the MC phase would be $500$ s$^{-1}$ which, indeed, is the slope of the $x-t$ straight line in the corresponding inset of Fig.\ref{fig-xt0koff}.

\subsection{Non-vanishing rate of detachment}

\subsubsection{Low detachment limit: heuristic analytical argument}

Let us consider the low, but non-vanishing, values of the rate of detachment $k_{off}$.

\subsubsection{High detachment limit: approximate theory based on first-passage analysis}

Next let us consider the opposite limit, namely, the high detachment limit.  In this case, the searcher TF dissociates even before it feels the strong directional push of the unidirectional flow of RNAPs. Therefore, in this limit the recently developed theoretical framework \cite{veksler13} is expected to provide a reasonable description of the search dynamics. For a self contained discussion, we first summarize the main steps of the calculations reported in ref.\cite{veksler13} before presenting the new analytical formulas that we use for our work. 

Following Veksler and Kolomeisky \cite{veksler13}, we define the probability $F_{n|m}(t)$ to reach the target on site $m$ {\it for the first time} at time $t$  if at $t = 0$ the TF was at the site $n (n = 0, 1, ..., L)$. The time evolution of these first-passage probabilities are governed by the backward master equations \cite{veksler13},
\begin{eqnarray}
\frac{dF_{n|m}(t)}{dt}&=&b\bigl[F_{n+1|m}(t)+F_{n-1|m}(t)\bigr]+k_{off} F_{0|m}(t)\nonumber\\
&-& (2b + k_{off} )F_{n|m}(t)  ~({\rm for}~ 2 \le n \le L-1), \nonumber \\
\end{eqnarray}
while for the two ends of the lattice at $n = 1$ and $n = L$ the equations are
\begin{eqnarray}
\frac{dF_{1|m}(t)}{dt}&=&bF_{2|m}(t)+k _{off}F_{0|m}(t)-(b+k_{off} )F_{1|m}(t) \nonumber \\
\end{eqnarray}
\begin{eqnarray}
\frac{dF_{L|m}(t)}{dt}&=&bF_{L-1|m}(t)+k _{off}F_{0|m}(t)-(b+k_{off} )F_{L|m}(t) \nonumber \\
\end{eqnarray} 
If the TF starts from the solution, i.e., $n = 0$ according to our notation, then the corresponding backward master equation is given by \cite{veksler13},
\begin{eqnarray}
\frac{dF_{0|m}(t)}{dt}&=&\frac{k_{on}}{L}\sum_{n=1}^{L}F_{n|m}(t)-k _{on}F_{0|m}(t).
\end{eqnarray}

These equations can be analyzed by introducing Laplace transformations of first-passage probability functions, $\tilde{\cal F}_{n|m}(s)= \int_{0}^{\infty} e^{-st} F_{n|m} (t) dt$. Then, backward master equations can be rewritten as a set of simpler algebraic expressions,
\begin{eqnarray}
(s+2b+k_{off} )\tilde{\cal F}_{n|m}(s)&=&b[\tilde{\cal F}_{n+1|m}(s)+\tilde{\cal F}_{n-1|m}(s)] \nonumber \\
&+&k_{off}\tilde{\cal F}_{0|m}(s) 
\end{eqnarray}
\begin{equation}
(s+b+k_{off} )\tilde{\cal F}_{1|m}(s) = b\tilde{\cal F}_{2|m}(s)+k_{off}\tilde{\cal F}_{0|m}(s) 
\end{equation}
\begin{equation}
(s+b+k_{off} )\tilde{\cal F}_{L|m}(s) = b\tilde{\cal F}_{L-1|m}(s)+k_{off}\tilde{\cal F}_{0|m}(s) 
\end{equation}
\begin{equation}
(s+k_{on})\tilde{\cal F}_{0|m}(s) = \frac{k_{on}}{L}\sum_{n=1}^{L}\tilde{\cal F}_{n|m}(s)
\end{equation}
These equations are solved by assuming that the general form of the solution is $\tilde{\cal F}_{n|m}(s) = Ay^{n} + B$, and using boundary and initial conditions it yields
\begin{eqnarray}
\tilde{\cal F}_{n|m}(s)&=&\frac{(1-B)(y^{n} +y^{-n})}{y^{m}+y^{-m}}+B
\end{eqnarray}
for $1 \le n \le m$, and
\begin{eqnarray}
\tilde{\cal F}_{n|m}(s)&=&\frac{(1-B)(y^{1+L-n} +y^{n-L-1})}{y^{1+L-m}+y^{m-L-1}}+B
\end{eqnarray}
for $m \le n \le L$. Here, parameters $y$ and $B$ are given by
\begin{eqnarray}
y&=&\frac{s+2b+k_{off}-\sqrt{(s+2b+k_{off})^{2} -4b^{2}}}{2b},\\
B&=&\frac{k_{off}\tilde{\cal F}_{0|m}(s)}{(k_{off}+s)}.\\\nonumber
\end{eqnarray}
One can also show that
\begin{eqnarray}
\tilde{\cal F}_{0|m}(s)&=&\frac{k_{on}(k_{off} + s)S(s)}{Ls(k_{off} + k_{on} + s) + k_{off}k_{on}S(s)}
\end{eqnarray}
where the new auxiliary function $S(s)$ is given by
\begin{eqnarray}
S(s)&=&\frac{y(1 + y)(y^{-L} - y^{L})}{(1-y)(y^{1-m}+y^{m} )(y^{m-L}+y^{1+L-m})}
\end{eqnarray}
More specifically, the first-passage times to reach the target located at the site $m$, starting from any other site $n$, on the lattice can be computed from
\begin{eqnarray}
T_{n|m}&=&-\frac{d}{ds}\tilde{\cal F}_{n|m}(s)\bigg|_{s=0}.
\end{eqnarray}

\begin{figure}[h!]
  (a)\\[0.02cm]
    \includegraphics[angle=0,width=0.85\columnwidth]{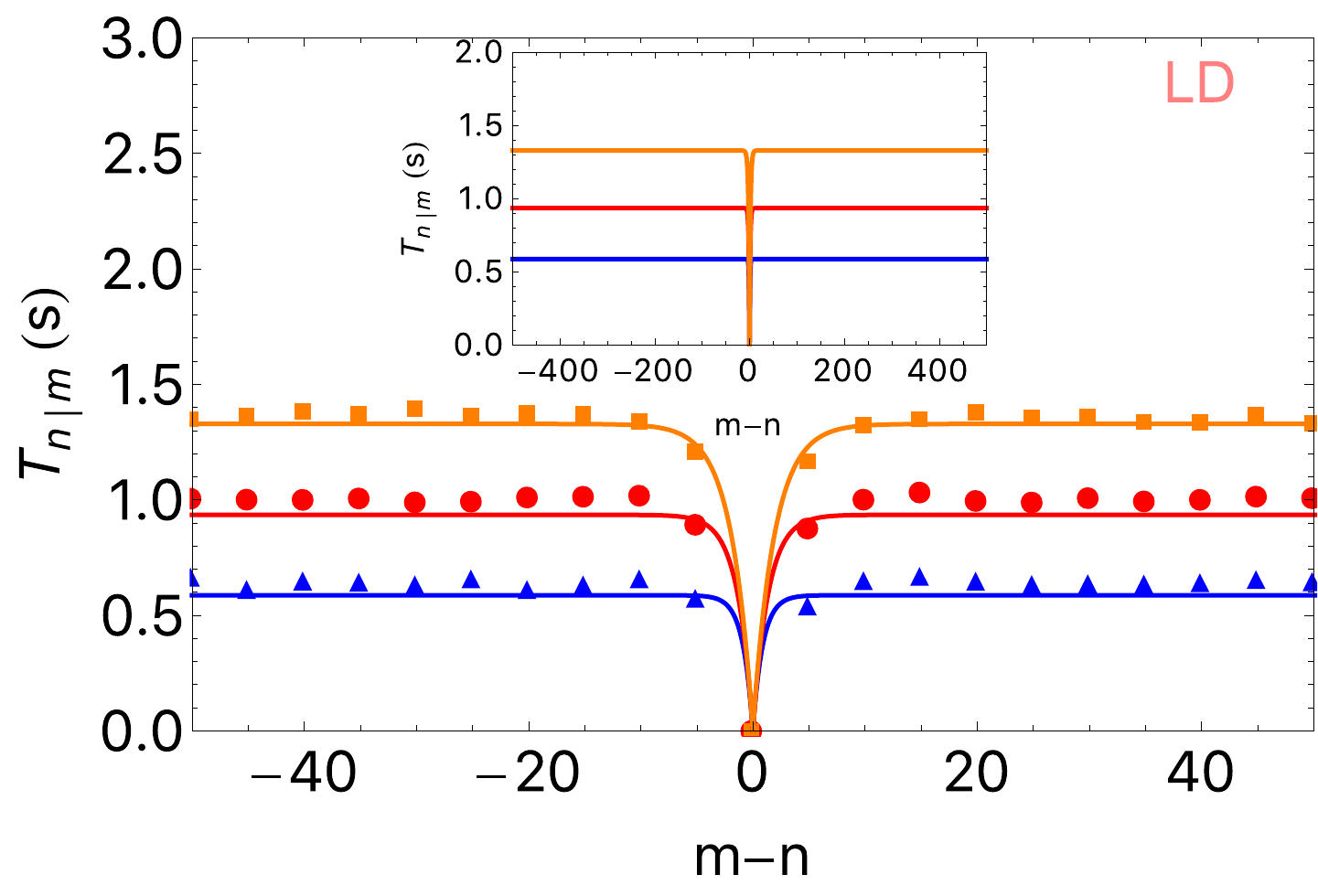} \\
  (b)\\[0.02cm]
    \includegraphics[angle=0,width=0.85\columnwidth]{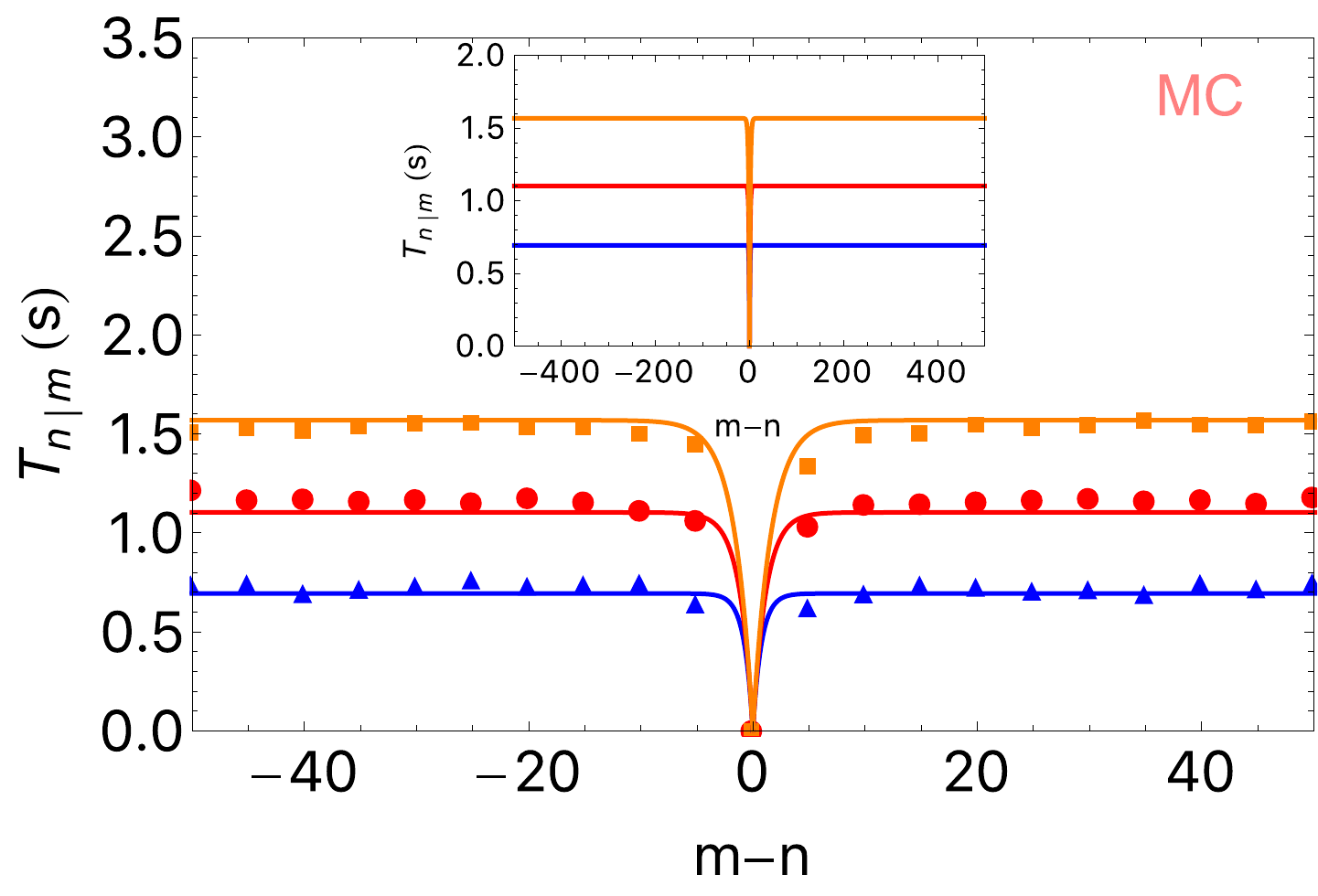} \\  
    (c)\\[0.02cm]
    \includegraphics[angle=0,width=0.85\columnwidth]{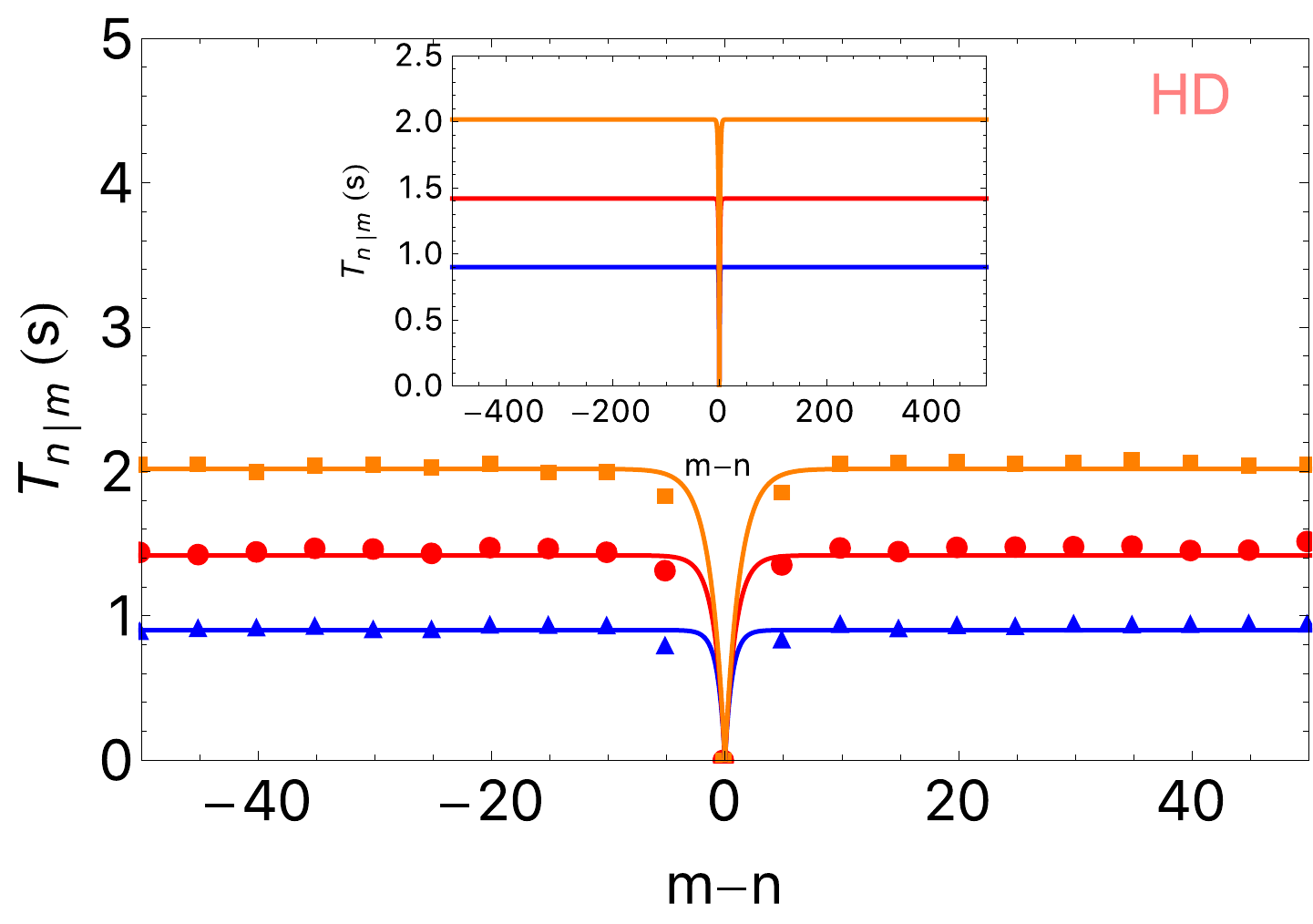} \\
  \caption{Variation of $T_{n|m}$ with respect to $m-n$ for different values of $k_{off}$, for (a) LD phase 
($\rho=0.3$), (b) MC phase ($\rho=0.5$) and (c) HD phase ($\rho=0.7)$. All other parameters are kept fixed at values : $b=10^{3}~ s^{-1}$, $k_{on}=10^{4}~ s^{-1}$, $m=L/2$ and $L=10^{3}$. Continuous lines have been obtained from extended Kolomeisky's formula and discrete data points have been obtained from MC simulation. Blue, red and orange color lines correspond to $k_{off}=1000~ s^{-1}$, $k_{off}=400~ s^{-1}$ and $k_{off}=200~ s^{-1}$ ,respectively. Triangle, circle and square correspond to $k_{off}=1000~ s^{-1}$, $k_{off}=400~ s^{-1}$ and $k_{off}=200~ s^{-1}$ ,respectively. The inset shows the variation of $T_{n|m}$ with respect to $m-n$ for different values of $k_{off}$ from $-L/2$ to $L/2$. Line colors are same for same parameters as before.}
  \label{fig-T_n_vs_n_Kolomeisky_theory}
\end{figure}

\begin{widetext}

\begin{eqnarray}
T_{n|m}&=&\biggl[2^{-1 - 2 m - n} \biggl(k_{on} + k_{off}\biggr) L x^{-1 - 2 m - n} \biggl(-k_{off} + \sqrt{k_{off} (k_{off} + 4 b)} \biggr)\nonumber\\
&&\biggl(-2^{m + 2 n} x^{m} + 2^{2 m + n} x^{n} + 2^{n} x^{2 m + n} - 2^{m} x^{m + 2 n}\biggr)\nonumber\\
&&\biggl\{-4^{m} k_{off} + 4^{m} \sqrt{k_{off} (k_{off} + 4 b)} - 2 b \biggl(4^{m} +x^{2 m}\biggr)\biggr\} \biggl\{-4^{m} k_{off} x^{2 L} \nonumber\\
&&+ 4^m \sqrt{k_{off} (k_{off} + 4 b)} x^{2 L}+2 b \biggl(4^{m} x^{2 L} + 4^{L} x^{2 m}\biggr)\biggr\}\biggr]\bigg/ \nonumber\\
&&\biggl[k_{on} k_{off} b^{2} \biggl\{k_{off} + 4 b - \sqrt{k_{off} (k_{off} + 4 b)}\biggr\} \biggl(4^{L} - x^{2 L}\biggr) \biggl(4^{m} + x^{2 m}\biggr)\biggr]~~~~\rm (for ~1 \le n \le m),
 \end{eqnarray} 
and
\begin{eqnarray}
T_{n|m}&=&\biggl[2^{-1 - 2 m - n} \biggl(k_{on} + k_{off}\biggr) L x^{-1 - 2 m - n} \biggl\{-k_{off} + \sqrt{k_{off} (k_{off} + 4 b)}\bigg\}\nonumber\\
&& \biggl(2^{n} x^{m} - 2^{m}x^{n}\biggr) \biggl\{-4^{m} k_{off} + 4^{m} \sqrt{k_{off} (k_{off} + 4 b)} - 2 b \biggl(4^{m} + x^{2 m}\biggr)\biggr\}\nonumber\\
&& \biggl\{-4^{m} k_{off} x^{2 L} + 4^{m} \sqrt{k_{off} (k_{off} + 4 b)} x^{2 L} - 2 b \biggl(4^{m} x^{2 L} + 4^{L} x^{2 m}\biggr)\biggr\}\nonumber\\
&&\biggl\{-2^{m + n}k_{off}^{2} x^{2 L} + 2^{m + n} k_{off} x^{2 L} \biggl(-4 b + \sqrt{k_{off} (k_{off} + 4 b)}\biggr) \nonumber\\
&&+ 2 b \biggl(-2^{m + n} b x^{2 L} +2^{m + n} \sqrt{k_{off} (k_{off} + 4 b)} x^{2 L} + 4^{L} b x^{m + n}\biggr)\biggr\}\biggr]\bigg/\nonumber\\
&&\biggl[k_{on} k_{off} b^{2} \biggl(k_{off} + 4 b - \sqrt{k_{off} (k_{off} + 4 b)}\biggr) \biggl(4^{L} - x^{2 L}\biggr) \biggl\{4^{m} k_{off}^{2} x^{2 L} - 4^{m} k_{off} x^{2 L}\nonumber\\ 
&&\biggl(-4 b + \sqrt{k_{off} (k_{off} + 4 b)}\biggr)+ 2 b \biggl(4^{m} b x^{2 L} - 4^{m} \sqrt{k_{off} (k_{off} + 4 b)} x^{2 L} + 4^{L} b x^{2 m}\biggr)\biggr\}\biggr]~\rm (for ~m \le n \le L)
\end{eqnarray}
\end{widetext}
where 
\begin{equation}
x=\biggl(k_{off}+2b-\sqrt{k_{off}(k_{off}+4b)}\biggr)\bigg/b. 
\end{equation}
The average time to find the target, starting from the solution, $T_{0}$, can be easily found using the following equality,
\begin{eqnarray}
T_{0}&=&-\frac{d}{ds}\tilde{\cal F}_{0|m}(s)\bigg|_{s=0}\nonumber\\
&=&\frac{k_{off} L+k_{on}(L-S(0))}{k_{on}k_{off}S(0)}.
\end{eqnarray}

The theoretical treatment summarized above does not include traffic-like flow of the RNAPs. In order to capture these effects, the diffusion rate $b$ and the attachment rate $k_{on}$ have been replaced by the effective rates $b(1-\rho)$ and $k_{on}(1-\rho)$, respectively, where $\rho$ is the steady state density of the RNAPs. Note that, in the absence of the TF, the density $\rho$ of the RNAPs is determined by the magnitudes of the three rates $\alpha, \beta$ and $u$.

In order to check how well this theory works in the limit of high detachment rates, we have carried out MC simulations. First, in the absence of any TF, the system of RNAPs are allowed to evolve for one million MC steps allowing their traffic flow to reach the steady value during that period. Then a TF is added to the solution (in the state $0$) and the composite system consisting of the TF and the RNAPs is allowed to evolve following the dynamical rules summarized by Fig.\ref{fig-model}. Each round of the search process ends as soon as the TF reaches the target site; the next round of search begins with a new TF in solution.  The time taken by the TF in each round of search is recorded and the data are finally averaged over 2000 rounds of search to calculate the mean first-passage time, i.e., the average search time, needed by a single TF searcher to reach the target. 

The numerical values of the rates $\alpha, \beta$ and $u$ were chosen in such a way that the traffic of the RNAPs, in the absence of the TF, would attain the desired dynamical phase. The particular sets of values chosen for the three phases LD, MC and HD phases are as follows: LD : $\alpha =300~s^{-1}, ~\beta=1000~s^{-1}$; for MC : $\alpha =1000~s^{-1}, ~\beta=1000~s^{-1}$ and for HD : $\alpha =1000~s^{-1}, ~\beta=300~s^{-1}$. We have used the value $u=1000~s^{-1}$ for all three phases. Corresponding to these values of the parameters, over the range of $1000~ s^{-1}  \geq k_{off} \geq 200~ s^{-1}$ the detachment rate $k_{off}$ is still sufficiently high so that excellent agreement between the theoretical predictions and the data obtained from MC simulation are seen in Fig.\ref{fig-T_n_vs_n_Kolomeisky_theory}. This level of agreement with the MC data established that the approximations made in the theoretical derivations are well justified in this regime.  Note that, because of the rapid detachments of the TF from the lattice in the parameter regime used for the plots in Fig.\ref{fig-T_n_vs_n_Kolomeisky_theory}, the duration of its each round of scanning is too short to be significantly affected by the flow of the RNAPs. Therefore, it is not surprising that the theory is in good agreement with the MC simulation data in this regime.


\begin{figure}[h!]
  (a)\\[0.02cm]
    \includegraphics[angle=0,width=0.85\columnwidth]{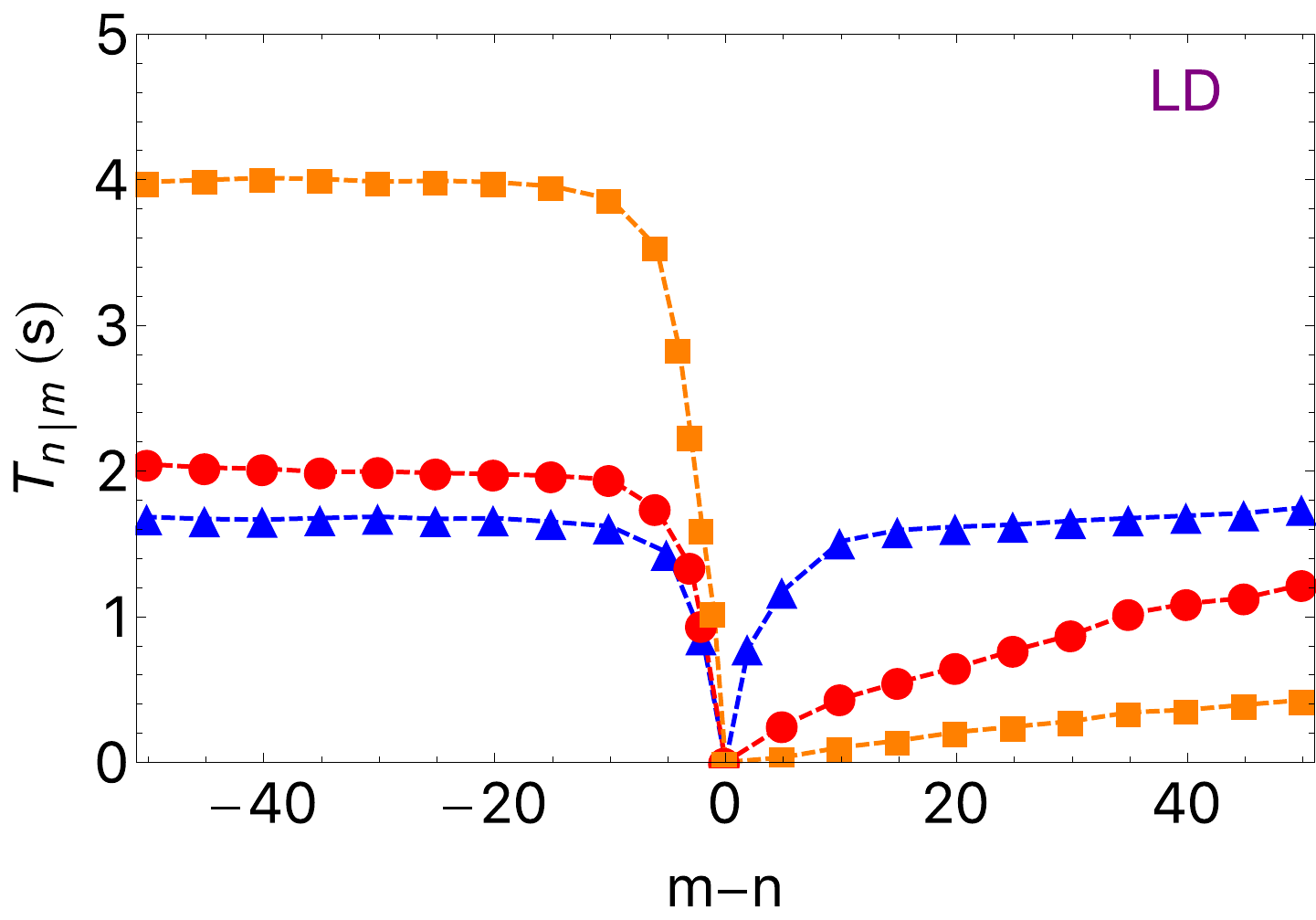} \\
  (b)\\[0.02cm]
    \includegraphics[angle=0,width=0.85\columnwidth]{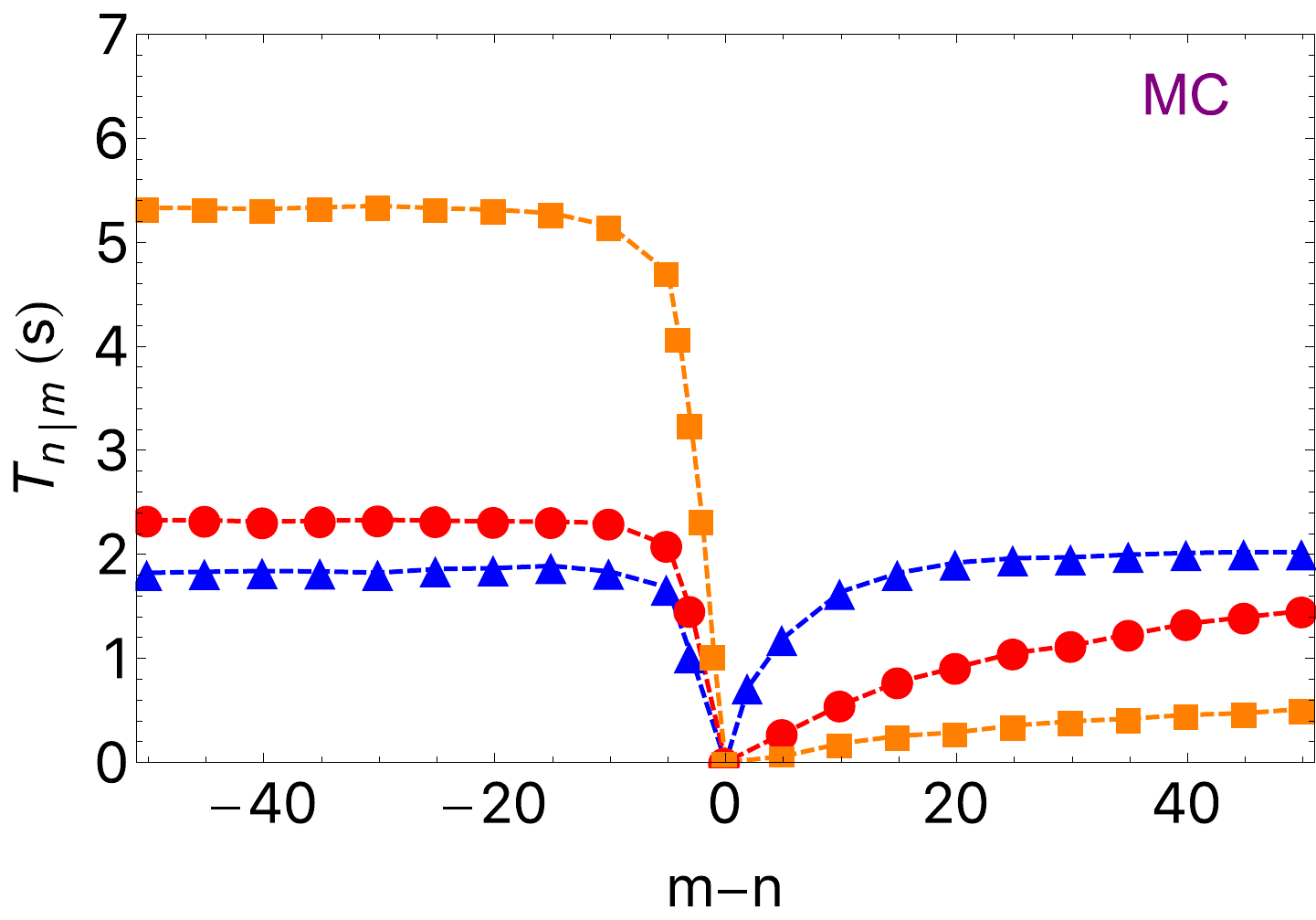} \\  
    (c)\\[0.02cm]
    \includegraphics[angle=0,width=0.85\columnwidth]{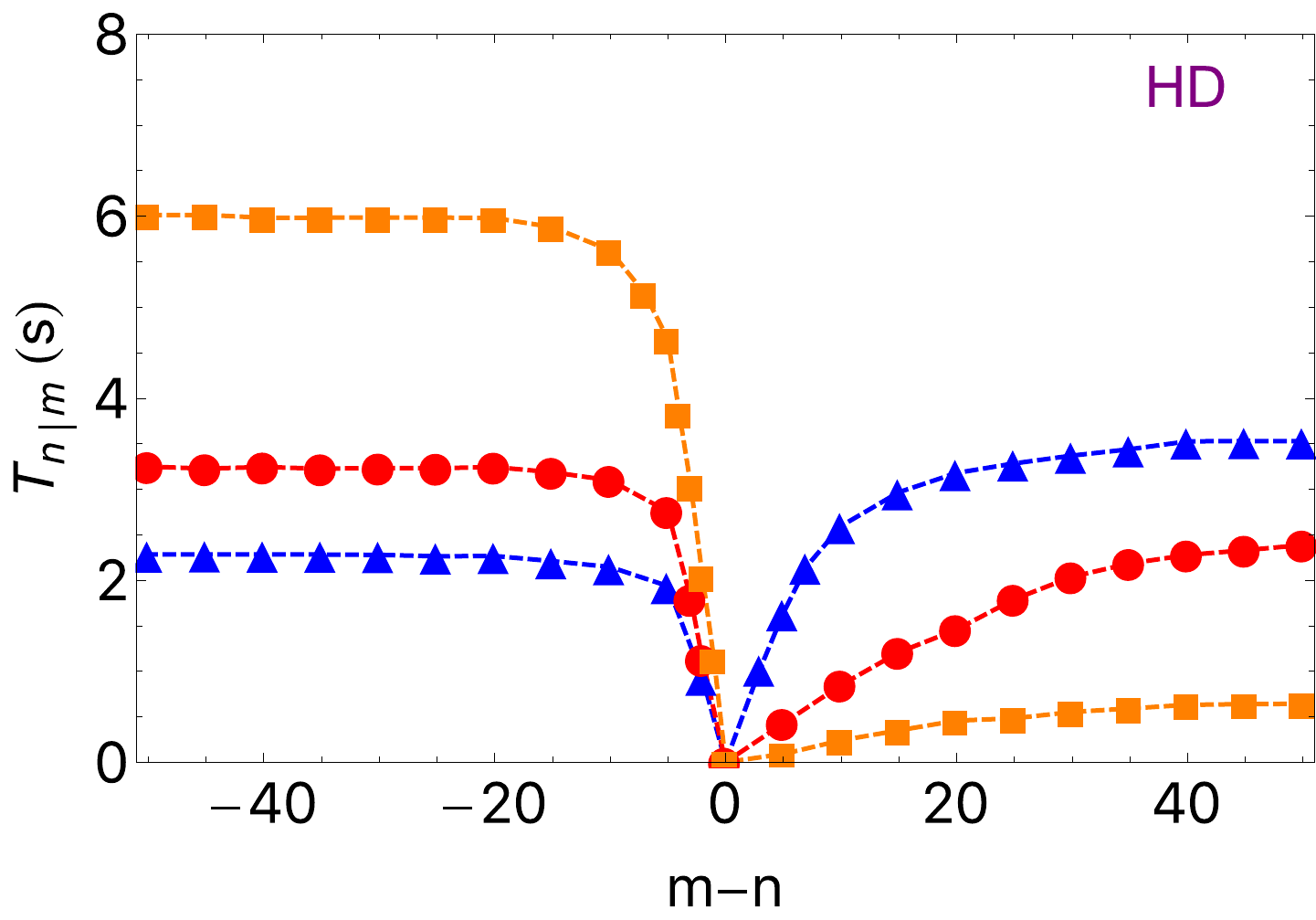} \\
  \caption{Variation of $T_{n|m}$ with respect to $m-n$ for different values of $k_{off}$, for (a) LD phase 
($\rho=0.3$), (b) MC phase ($\rho=0.5$) and (c) HD phase ($\rho=0.7$). All the other parameters are kept fixed at values : $b=10^{3}~ s^{-1}$, $k_{on}=10^{4}~ s^{-1}$, $m=L/2$ and $L=10^{3}$. The discrete data points have been obtained from MC simulation. Blue triangle, red circle and orange square correspond to $k_{off}=100~ s^{-1}$, $k_{off}=10~ s^{-1}$ and $k_{off}=1~ s^{-1}$, respectively. Dashed lines are drawn just as guide to the eyes.}
  \label{fig-T_n_vs_n_Kolomeisky_theory2}
\end{figure}


But, as the detachment rate $k_{off}$ decreases, the TF spends longer times scanning the lattice, and most of its attempts to step against the flow of the RNAPs become unsuccessful in comparison with its steps in the direction of flow. The resulting motion of the TF is similar to that of a Brownian ratchet \cite{reimann02,julicher97}. The curves showing $T_{n|m}$ are now asymmetric about $m-n=0$; the lower is the value of $k_{off}$ the stronger is this asymmetry (see Fig.\ref{fig-T_n_vs_n_Kolomeisky_theory2}). Since in this regime the Veksler-Kolomeisky theory \cite{veksler13} shows large deviation from the MC data and the deviation increases with decreasing $k_{off}$ we have plotted only the MC data in Fig.\ref{fig-T_n_vs_n_Kolomeisky_theory2}; the lines connecting these discrete data points serves merely as a guide to the eye.  

When the TF is attached to the lattice it scans a distance of the order $2 \lambda$, where $\lambda=\sqrt{b_{eff}/k_{off}}$ during each encounter that lasts for a time of the order of $1/k_{off}$. Here $b_{eff}$ is the effective diffusion rate of the TF. If we take the smallest $k_{off} = 1$s$^{-1}$ and the largest $b_{eff}=1000$s$^{-1}$ the scanning distance is of the order of $70$.  For $k_{off}=10$ s$^{-1}$ it will be about $27$, while for $k_{off}=100$ s$^{-1}$ it will be about $7$. One would see the asymmetry displayed in 
Fig.\ref{fig-T_n_vs_n_Kolomeisky_theory2}  if the TF are not farther than this distance from the target, irrespective of whether it is located to the left or right of the target. Because, on the average, the TF downstream will have difficulty to find the target (by colliding with the moving RNAPs) and it will have to dissociate, while the TF upstream can find it without dissociation.   Let us call the distance $2\lambda$ upstream and $2\lambda$ downstream as 2 `��antenna' zones.

It is easy to explain the observed asymmetry almost quantitatively. Suppose the TF starts in the middle of the antenna zone upstream, say with initial $m-n=35$. Then the search time can be estimated as $T=1/2k_{off}$. The coefficient $1/2$ appears because the TF starts only in the middle of the antenna region. If we take $k_{off}=1$ s$^{-1}$, then we get the search time to be about $0.5$ s. This is exactly what we see in Fig. 4. Now if the TF starts in the middle of the antenna zone downstream (i.e., say,with initial $m-n=-35$) then the searcher cannot reach the target because of the oncoming traffic of RNAPs unless it dissociates. It will spend $0.5$  s here and then, after dissociation, it will have the chance to bind to the upstream region and do better. But, on the average, it will come to the position $L/4$ from the target, which is in the middle of the left half of the chain. On the  average, it will have to do $(L/4)/(2\lambda)$ cycles, each of approximate average duration $1/k_{off}$  (because $k_{on}=10000$ s$^{-1}$ $>>1$). For $k_{off}=1$ s$^{-1}$ the searcher will need to make $3.5$ cycles, on the average. So, the total search time for a TF starting in the downstream antenna region is $0.5s+3.5s=4$ s  which is in excellent agreement with the corresponding simulation data plotted in Fig. \ref{fig-T_n_vs_n_Kolomeisky_theory2} for the LD phase. It will take longer in MC and HD phases because the effective diffusion rates here are smaller due to the smaller average spacings between the RNAPs. So it takes a little bit longer. Because $\rho_{HD}=0.7$ and $\rho_{LD}=0.3$ it would take longer $\sqrt{\rho_{HD}/\rho_{LD}}=1.5$, or $T=6$ s. This is also in excellent agreement with the simulation data. 

\begin{figure}[th]
(a)\\[0.02cm]
    \includegraphics[angle=0,width=0.9\columnwidth]{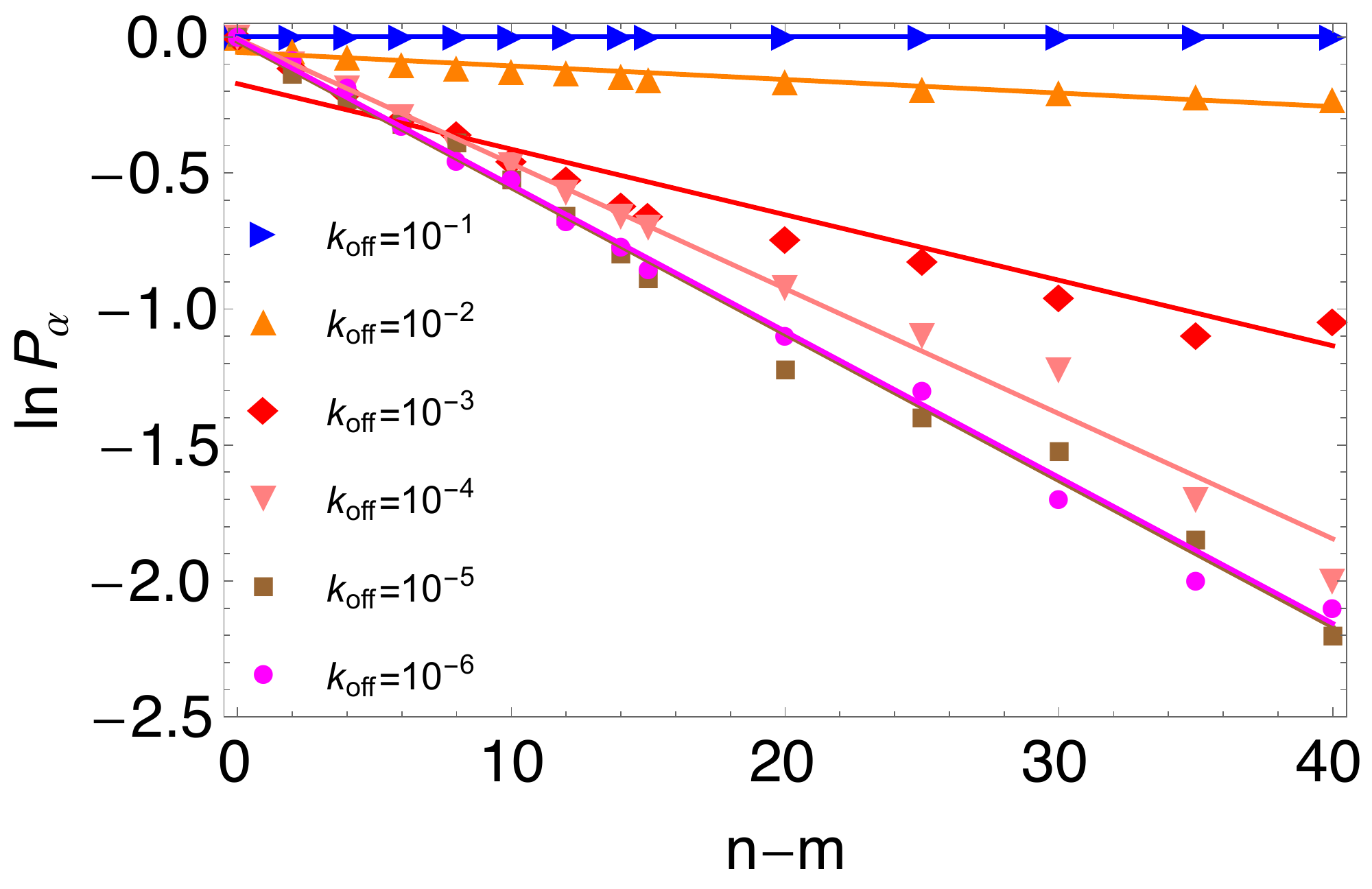} \\[0.02cm]
(b)\\[0.02cm]    
     \includegraphics[angle=0,width=0.9\columnwidth]{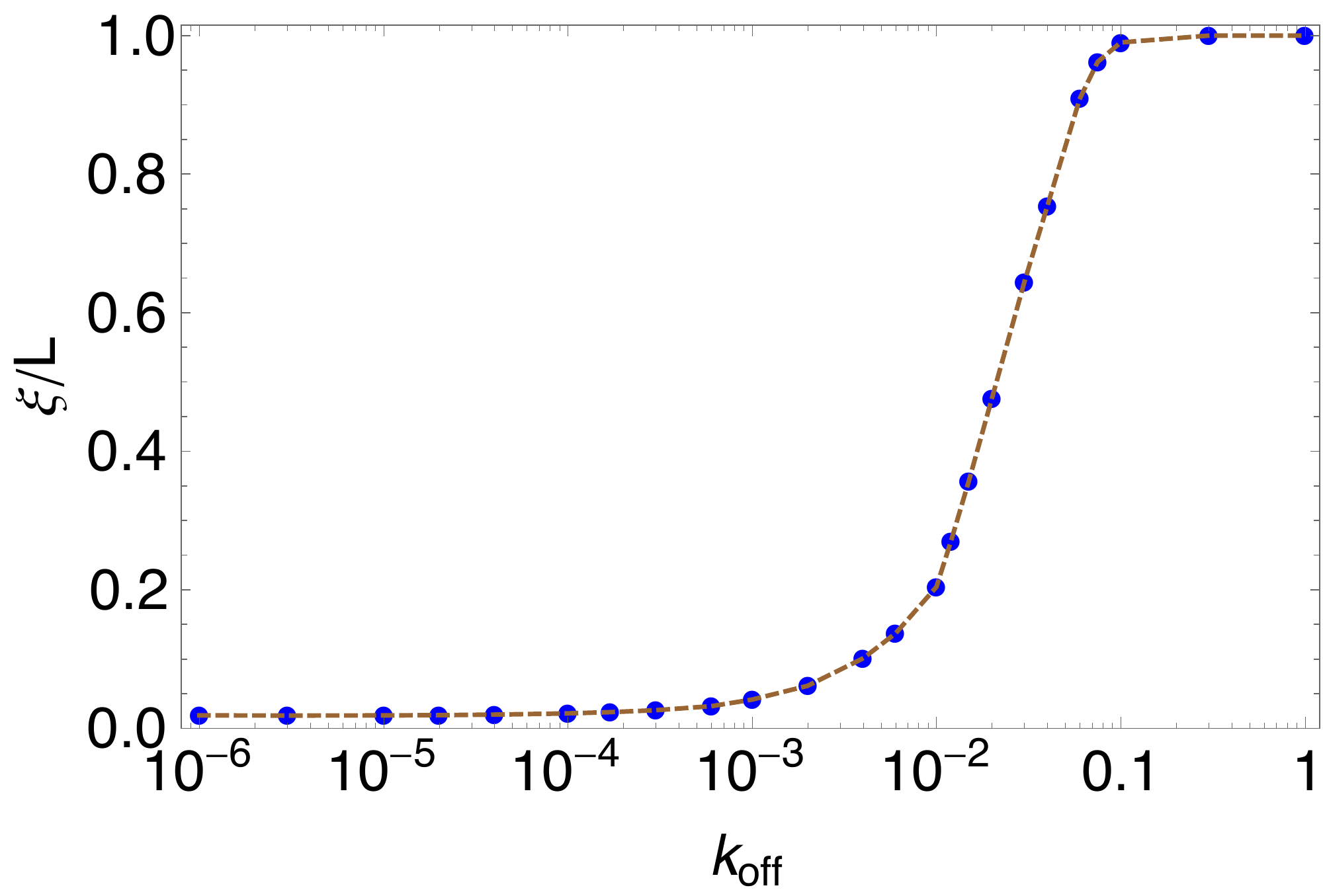} \\
    \caption{(a) LD ($\rho=0.01$) : Logarithm of the probability of the successful search events {\it in finite time} are plotted with respect to relative distance of initial position of TF from the target. Data points are obtained from MC  simulation. Lines correspond to best fit curves. Different lines are for different values of $k_{off}$.  (b) Inverse of slopes (correlation lengths) of these lines are plotted against $k_{off}$. Dashed line is drawn just to guide the eyes.}
  \label{fig-prob}
\end{figure}

In reality, search time available to a TF is finite. What is the effect of such {\it finite time of search} on the success in hitting the target site? 
If a TF begins search from a site located downstream from the target site, it cannot reach the target site if $k_{off}=0$. In other words, the probability of hitting the target is zero for all $n > m$ in the special case $k_{off}=0$. Even for small values of $k_{off}$, only a fraction of the search attempts are successful is reaching the target site {\it over a finite duration of search}. In our MC simulations, for each given initial value of $|n-m|$, we define the probability of successful events ($P_{\alpha}$) as the number of successful events divided by the total number of events when each unsuccessful search attempt is aborted after $10^6$ MC steps. The parameter values chosen for this study are 
$\alpha =10~s^{-1}, ~\beta=1000~s^{-1}$. $u=1000~s^{-1}$ which, in the absence of any TF would lead to the LD phase of the RNAPs in the steady state. All other parameters are kept fixed at values : $b=10^{3}~ s^{-1}$, $k_{on}=10^{6}~ s^{-1}$, $m=L/2$ and $L=10^{3}$. In Fig.\ref{fig-prob} we have computed the logarithm of the probability of the successful events as a function of $|n-m|$ for several different values of the parameter $k_{off}$. The data fit well with straight lines indicating that 
\begin{equation}
P_{\alpha}  \propto exp(-|n-m|/\xi)
\end{equation}
where the normalized range $\xi/L$ of successful search increases from 0 to 1 with the increase of $k_{off}$.


\subsection{Mean search time starting from solution}

Let the symbols $b_{eff}$ and $v_{eff}$ denote the effective velocities of TF and RNAPs, respectively. Similarly, $(k_{on})_{eff}$ is the effective attachment rate of the TF on the track per site. $k_{off}$ is the detachment rate of the TF from any site on track. Suppose $\rho$ is the steady state number density of the RNAPs while $\alpha$ and $\beta$ denote the rates of their attachment and detachment, respectively, with the track. 
From known TASEP results we can write,
\begin{eqnarray}
\rho = \alpha/u {\rm~in ~LD}\\
b_{eff} = b(1-\rho)\\
v_{eff} = u(1-\rho)\\
(k_{on})_{eff}=k_{on}(1-\rho)
\end{eqnarray}
where $b$ and $u$ are the hoping rates of TF and RNAPs, respectively and $k_{on}$ is the attachment rate of the TF on the track per site.  
Since the oncoming traffic of RNAPs reduce the effective velocity of the TF, the effective speed of the TF is reduced to $b_{eff}-v_{eff}$ when it moves against flow of the RNAPs. In contrast, the TF can hop in the direction of flow of the RNAPs with the effective speed $b_{eff}$ as, in this case, except for the fact that $b_{eff} \neq v_{eff}$, the TF and RNAPs move in same direction respecting the same exclusion principle. 

We use the integer indices $n=1,2,3,.......,L$ to label the equispaced sites on the track, where the integers increase in the direction of movement of the RNAPs. Let $P(n,t)$ denote the probability that there is a TF at site $n$ at time $t$ where $n=0$ represents solution.The stochastic time evolution of the system, in terms of these probabilities, is governed by the master equations
\begin{widetext}
\begin{eqnarray}
\frac{dP(0,t)}{dt}&=&k_{off}\sum_{n=1}^{L/2-1}P(n,t)+k_{off}\sum_{n=L/2+1}^{L}P(n,t)- L~(k_{on})_{eff}~P(0,t)~ ,\\ 
\frac{dP(1,t)}{dt}&=&\bigl(b_{eff}-v_{eff}\bigr)~P(2,t)+(k_{on})_{eff}~P(0,t)-\bigl(b_{eff}+k_{off}\bigr)~P(1,t)~ ,\\ 
\frac{dP(L/2-1,t)}{dt}&=&b_{eff}~P(L/2-2,t)+(k_{on})_{eff}~P(0,t)-\bigl(2~b_{eff}-v_{eff}+k_{off}\bigr)~P(L/2-1,t)~ ,\\ 
\frac{dP(L/2,t)}{dt}&=&b_{eff}~P(L/2-1,t)+\bigl(b_{eff}-v_{eff}\bigr)~P(L/2+1,t)+(k_{on})_{eff}~P(0,t)~ ,\\
\frac{dP(L/2+1,t)}{dt}&=&\bigl(b_{eff}-v_{eff}\bigr)~P(L/2+2,t)+(k_{on})_{eff}~P(0,t)-\bigl(2~b_{eff}-v_{eff}+k_{off}\bigr)~P(L/2+1,t)~ ,\\ 
\frac{dP(L,t)}{dt}&=&b_{eff}~P(L-1,t)+(k_{on})_{eff}~P(0,t)-\bigl(b_{eff}-v_{eff}+k_{off}\bigr)~P(L,t)~ ,\\
\frac{dP(n,t)}{dt}&=&b_{eff}~P(n-1,t)+\bigl(b_{eff}-v_{eff}\bigr)~P(n+1,t)+(k_{on})_{eff}~P(0,t)\nonumber\\
&-&\bigl(2~b_{eff}-v_{eff}+k_{off}\bigr)~P(n,t)~~~~~({\rm for}~ 1<n<L/2-1~ {\rm and}~L/2+1<n<L)~ .
\end{eqnarray}
By solving these master equations iteratively, starting with the initial condition $P(n=0,t=0)=1$, i.e; with the TF initially in the solution, we can get the above probabilities for all subsequent instants of time $t > 0$. Utilizing these solutions, we get the probability density of the search times. More specifically, the probability density $f(t)$ to reach the target site located at $m=L/2$ between the times $t$ and $t+dt$ is obtained from
\begin{eqnarray}
f(t)&=&b_{eff}~P(L/2-1,t)+\bigl(b_{eff}-v_{eff}\bigr)~P(L/2+1,t)+(k_{on})_{eff}~P(0,t)
\end{eqnarray}

In the actual numerical calculation, we stop the process of iterative solution of the set of coupled master equations at a time $t=T_{max}$ when $P(m=L/2,t \to \infty)$ becomes $\approx$ 1. The mean search time for a TF, defined by  
\begin{eqnarray}
<T_{s}>&=&\int_{0}^{\infty}t~f(t) dt
\end{eqnarray}
\end{widetext}
is then numerically computed by evaluating the integral after replacing the upper limit of the integral by $T_{max}$. 

In our computation of the mean search time by MC simulation, we start with empty lattice and switch on the flow of RNAPs. We monitor the flow of the RNAPs for the first one million time steps to ensure that the system reaches steady state. Then an absorbing boundary condition for the TF is imposed at the designated target site at $m=L/2$. We place a TF in the solution ($n=0$) and start our clock ($t=0$). The updating of the state of the combined system of RNAPs and TF is continued till the TF reaches the target site; at that instant the simulation is stopped and the corresponding clock reading gives the search time for that particular MC run. We repeat this same procedure for 1000 MC runs and then average over all the MC runs to calculate the mean search time $(<T_{s}>)$ for TF to find the target located at $m=L/2$. 

\begin{figure}[h]
  \includegraphics[angle=0,width=0.75\columnwidth]{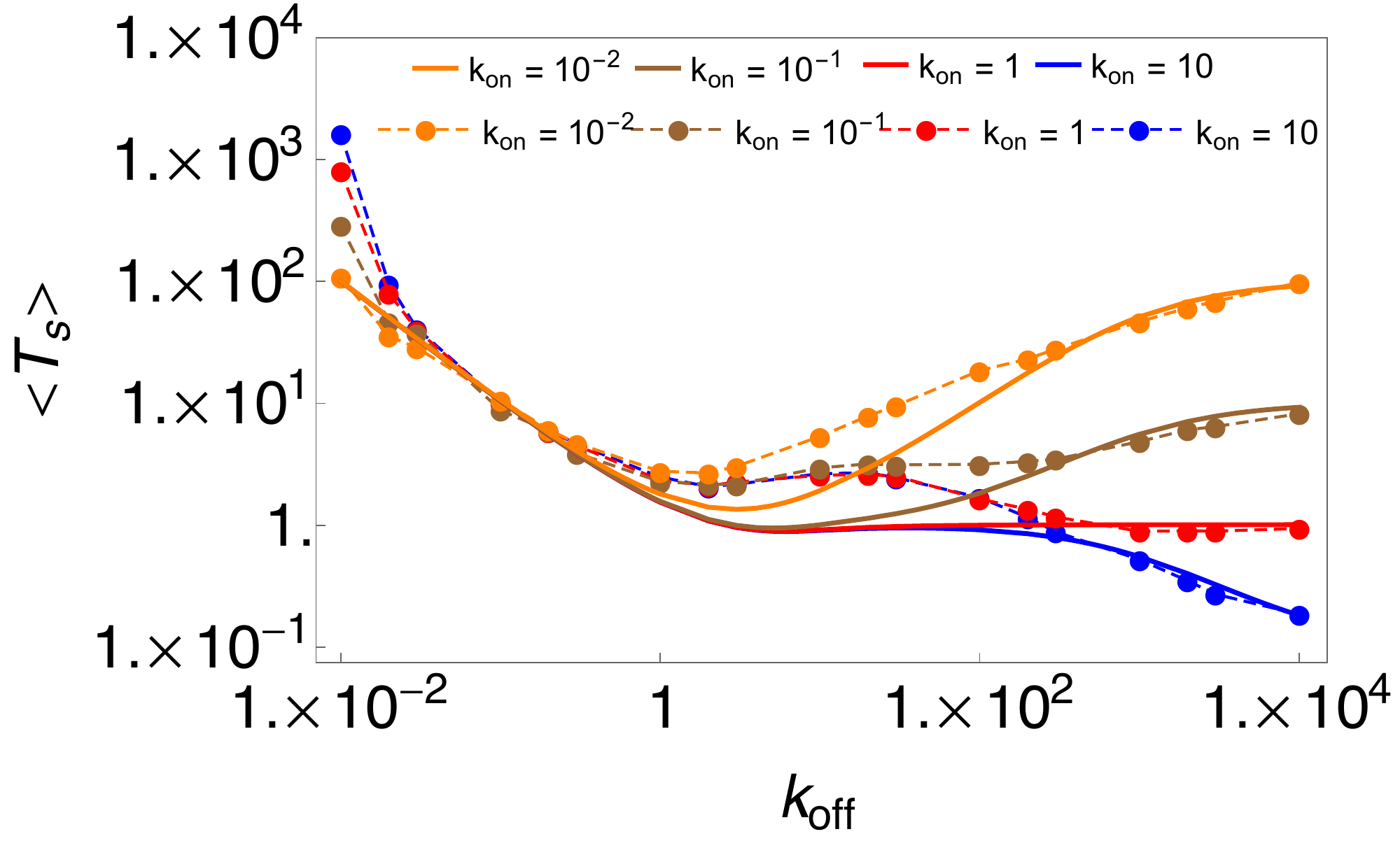}
  \caption{LD: Mean search time is plotted against $k_{off}$ for four different values of $k_{on}$. Parameters are kept fixed at values : $\alpha =10~s^{-1}$, $\beta=10^{3}~s^{-1}$, $u=10^{3}~s^{-1}$, $\rho=0.01$, $b=10^{3}~ s^{-1}$, $dt=5 \times 10^{-5}~s$, $m=L/2$ and $L=10^{3}$. Initially TF was in solution $(n=0)$. Continuous lines correspond to mean-field (MF) theory. Discrete data points are obtained from simulation; $T_{s}$ has been averaged over 1000 MC runs to get $<T_{s}>$. Dashed lines are drawn just to guide the eyes. }
  \label{fig-5}
\end{figure}

The results of the numerical solutions of the master equations and those obtained from MC simulations are plotted in Fig.\ref{fig-5}. The predictions of the MFA are in good agreement with the MC data. An interesting feature of the $<T_{s}>$ -vs-$k_{off}$ curves is the occurrence of a (local-) minimum indicating an optimal search strategy. Moreover, for a given $k_{on}$, the search can be made even more efficient than that implied by the local minimum by tuning $k_{off}$ to an appropriately large value.

\section{Summary and Conclusion}

In this paper we have developed a kinetic model for the search of a specific binding site on a linear chain by a single particle that executes diffusive motion along the chain in the presence of a uni-directional traffic flow of another distinct species of particles. This phenomenon resembles diffusive search conducted by a protein, called transcription factor (TF), for its specific binding site on a DNA while a stream of RNA polymerase (RNAP) motors move collectively in a uni-directional traffic-like manner on the same segment of DNA. At first sight, one might expect that such a huge crowd of RNAP might cause strong hindrance against the natural movement of the TF thereby increasing the time it requires for hitting the target site on the DNA. Contrary to this naive expectation, we find that, over wide range of values of the kinetic parameters of this model, the search required shorter time because the RNAPs can reduce wasteful excursions in the wrong directions by pushing the TF in the correct direction. More precisely, the Brownian diffusion of the TF gets rectified to a pattern of movement that can be identified with a Brownian ratchet \cite{reimann02,julicher97}. Thus, in the presence of the RNAP traffic, the mean time of successful search can be even shorter than that required in the absence of RNAP traffic.  

Once a TF detaches from the DNA, it can resume its diffusive search only after re-attaching to the DNA. However, if  the site of re-attachment is random, the distance between the site of its re-attachment and the target site may be longer than that between its location just before detachment and the target site. Thus, the TF does not draw any benefit from its earlier search history. So, detachment may appear to disrupt the search process. But, that is not true,    
 as we show in this paper. When the TF finds it practically impossible to move towards the target site by hopping against the flow of RNAP traffic, detachment from the DNA gives it a fresh opportunity re-start search from another location from where it can move co-directionally with the RNAP traffic. In fact, in the latter situation, instead of hindering the search by the TF, the RNAPs  assist its search by dragging it downstream along with them. However, too frequent detachment can be detrimental for the successful completion of the search process. Based on these intuitive arguments, one would expect an optimal rate of detachment that would correspond to fastest search, i.e., the shortest mean search time. This is exactly what we show analyzing our kinetic model analytically under mean-field approximation as well as by direct MC simulation. 

The model developed here does not include the non-motor crowders on the lattice. Therefore, it does not account for the effects of DNA-bound proteins like histones, etc. on the time of search by the TF. Moreover, effects of elastic forces arising from bending and possible twisting of DNA are not not incorporated in our calculations. We hope to extend our model in future including both these features to describe the search by TF more realistically. Nevertheless, we hope, our work would motivate test of the validity of the theoretically predicted phenomena mentioned above by carrying out experiments {\it in-vitro} using a single DNA strand stretched by applying tension at its two ends with optical tweezers.

\section*{Acknowledgments}
 Work of one of the authors (BM) has been supported by a Senior Research Fellowship from UGC. ABK acknowledges the  support from the Welch Foundation (Grant No. C-1559), from the NSF (Grant No. CHE-1664218), and from the Center for Theoretical Biological Physics sponsored by the NSF (Grant No. PHY-1427654). DC acknowledges support from SERB through a J.C. Bose National Fellowship. The authors thank ICTS for the hospitality in Bangalore where this work was initiated during the ICTS program ``Collective Dynamics of-,on- and around Filaments in Living Cells: Motors, MAPs, TIPs and Tracks''.


\end{document}